\documentclass[11pt]{article}
\def\beqa{\begin{eqnarray}}
\def\eeqa{\end{eqnarray}}

\usepackage{graphics}
\begin{document}
\begin{titlepage}
         \title{Radiation Bursts from Particles in the Field of Compact,
         Impenetrable, Astrophysical Objects}
\author{G. Papini$^{a, b}$\thanks{E-mail: papini@uregina.ca},
 G. Scarpetta$^{b,c,d}$\thanks{E-mail: scarpetta@sa.infn.it},
 V. Bozza$^{c,d}$\thanks{E-mail: bozza@sa.infn.it},
 A. Feoli$^{d,e}$\thanks{E-mail: feoli@unisannio.it}. \\
 G. Lambiase$^{c,d}$\thanks{E-mail: lambiase@sa.infn.it}.  \\
 {\em $^a$Department of Physics, University of Regina,} \\
 {\em Regina, Sask. S4S 0A2, Canada.} \\
 {\em $^b$International Institute for Advanced Scientific Studies,} \\
 {\em  84019 Vietri sul Mare (SA), Italy.} \\
 {\em $^c$Dipartimento di Fisica ''E.R.Caianiello'', Universit\'a di Salerno} \\
   {\em  84081 - Baronissi (SA), Italy} \\
 {\em $^d$INFN, Gruppo collegato di Salerno, Italy.} \\
 {\em $^e$Facolt\'a d'Ingegneria, Universit\'a del Sannio, 82100 Benevento, Italy.} \\
 }
              \date{\today}
              \maketitle

              \begin{abstract}
The radiation emitted by charged, scalar particles in a Schwarzschild field with maximal
acceleration corrections is calculated classically and in the tree approximation of
quantum field theory. In both instances the particles emit radiation that has
characteristics similar to those of gamma-ray bursters.
 \end{abstract}
\thispagestyle{empty} \vspace{20. mm}

              \vfill
          \end{titlepage}

Compact, impenetrable, astrophysical objects (CIAOs)\cite{sch,boson,reiss,kerr} arise in
a model, proposed by Caianiello and collaborators\cite{caia}, in which particle
accelerations have an upper limit ${\cal A}_m=2mc^3/\hbar$, referred to as maximal
acceleration (MA). The limit can be derived from quantum mechanical considerations
\cite{ca,pw}, is a basic physical property of all massive particles and must therefore be
included in the physical laws from the outset. This requires a modification of the metric
structure of space-time.

Classical and quantum arguments supporting the existence of a MA
have been discussed in the literature \cite{prove,wh,b}. MA also
appears in the context of Weyl space \cite{pap}, and of a
geometrical analogue of Vigier's stochastic theory \cite{jv} and
plays a role in several issues. It is invoked as a tool to rid
black hole entropy of ultraviolet divergences \cite{McG} and of
inconsistencies arising from the application of the point-like
concept to relativistic particles \cite{he}. MA may be also
regarded as a regularization procedure \cite{nesterenko} that
avoids the introduction of a fundamental length \cite{gs}, thus
preserving the continuity of space-time.

An upper limit on the acceleration also exists in string theory where Jeans-like
instabilities occur \cite{gsv,gasp} when the acceleration induced by the background
gravitational field reaches the critical value $a_c = \lambda^{-1} = (m\alpha)^{-1}$
where $\lambda$, $m$ and $\alpha^{-1}$ are string size, mass and tension. At
accelerations larger than $a_c$ the string extremities become casually disconnected.
Frolov and Sanchez \cite{fs} have also found that a universal critical acceleration must
be a general property of strings. It is the same cut--off required by Sanchez in order to
regularize the entropy and the free energy of quantum strings \cite{sa2}.

Applications of Caianiello's model include cosmology \cite{infl},
the dynamics of accelerated strings \cite{Feo}, the energy
spectrum of a uniformly accelerated particle \cite{emb}, neutrino
oscillations \cite{8,qua}, and photons in a cavity resonator
\cite{15}. The model also makes the metric observer--dependent,
as conjectured by Gibbons and Hawking \cite{Haw}.

The consequences of the model for the classical electrodynamics of
a particle \cite{cla}, the mass of the Higgs boson \cite{Higgs}
and the Lamb shift in hydrogenic atoms \cite{lamb} have been
worked out. In the last instance the agreement between
experimental data and MA corrections is very good for $H$ and $D$.
For $He^+$ the agreement between theory and experiment is improved
by $50\%$ when MA corrections are considered.

MA effects in muonic atoms appear to be measurable \cite{muo}. MA
also affects the helicity and chirality of particles \cite{chen}.

More recently, the model has been applied to particles falling in the gravitational field
of a spherically symmetric collapsing object \cite{sch}. In this problem MA manifests
itself through a spherical shell external to the Schwarzschild horizon and impenetrable
to classical and quantum particles \cite{boson}. The shell is not a sheer product of the
coordinate system, but survives, for instance, in isotropic coordinates. It is also
present in the Reissner-Nordstr\"om \cite{reiss} and Kerr \cite{kerr} cases. The model
therefore seems to preclude the usual process of formation of black holes. These are
replaced by CIAOs. The purpose of this work is to study the emission of radiation by a
scalar particles in the field of a CIAO.

Caianiello's model is based on an embedding procedure \cite{sch}
that stipulates that the line element experienced by an
accelerating particle is represented by
\begin{equation} \label{eq1}
d\tau^2=\left(1+\frac{g_{\mu\nu}\ddot{x}^{\mu}\ddot{x}^{\nu}}{{\cal
A}_m^2} \right)g_{\alpha\beta}dx^{\alpha}dx^{\beta}\equiv
\sigma^2(x) g_{\alpha\beta}dx^{\alpha}dx^{\beta}\equiv {\tilde g
}_{\alpha\beta}dx^\alpha dx^\beta\,,
\end{equation}
where $g_{\alpha\beta}$ is a background gravitational field. The effective space-time
geometry given by (\ref{eq1}) therefore exhibits mass-dependent corrections that in
general induce curvature and violations of the equivalence principle. The
four--acceleration $\ddot x^\mu = d^2 x^\mu/d\,s^2$ appearing in (\ref{eq1}) is a
rigorously covariant quantity only for linear coordinate transformations. Its
transformation properties are however known and allow the exchange of information among
observers. Lack of covariance for $\ddot x^\mu$ in $\sigma^2(x)$ is not therefore fatal
in the model. The justification for this choice lies primarily with the quantum
mechanical derivation of MA which applies to $\ddot x^\mu $, requires the notion of
force, is therefore Newtonian in spirit and is fully compatible with special relativity.
The choice of $\ddot x^\mu $ in (\ref{eq1}) is, of course, supported by the weak field
approximation to $g_{\mu\nu}$ which is, to first order, Minkowskian. On the other hand,
Einstein's equivalence principle does not carry through to the quantum level readily
\cite{lamm,singh}, and the same may be expected of its consequences, like the principle
of general covariance \cite{wein}. Complete covariance is, of course, restored in the
limit $\hbar \to 0$, whereby all quantum corrections, including those due to MA, vanish.

The corrections to the Schwarzschild field experienced by a
particle initially at infinity and falling toward the origin along
a geodesic, require that $\sigma^2(x)$ be calculated. From
(\ref{eq1}) and $\theta=\pi/2$, one finds (in units
$\hbar$=$c$=$G=1$)
 $$
\sigma^2(r)=1+\frac{1}{{\cal A}_m^2}\left\{
-\frac{1}{1-2M/r}\left(-\frac{3M\tilde{L}^2}{r^4}+\frac{\tilde{L}^2}{r^3}
-\frac{M}{r^2}\right)^2 + \right.
 $$
\begin{equation}\label{eq6}
\left. +\left(-\frac{4\tilde{L}^2}{r^4}+\frac{4\tilde{E}^2 M^2}{r^4(1-2M/r)^3}
\right)\left[\tilde{E}^2-\left(1-\frac{2M}{r}\right)\left(1+
\frac{\tilde{L}^2}{r^2}\right)\right]\right\}\,{,}
\end{equation}
where $M$ is the mass of the source, and $\tilde{E}$ and $\tilde{L}$ are the total energy
and angular momentum per unit of  particle mass $m$ \cite{sch}.

It is now convenient to introduce the adimensional variable
$\rho=r/M$ and the parameters $\epsilon=(M{\cal A}_m)^{-1}$ and
$\lambda={\tilde L}/M$. The analysis of the motion can be carried
out in terms of
\begin{equation}\label{eq11}
\tilde{V}^2_{eff}(\rho)=\tilde{E}^2\left\{1+\frac{1}{\sigma^2(\rho)}
\left[-\frac{1}{\sigma^2(\rho)}+\frac{1}{\tilde{E}^2}\left(1-\frac{2}{\rho}\right)\left(
1+\frac{\lambda^2}{\rho^2\sigma^2(\rho)}\right)\right]\right\}\,{.}
\end{equation}
Notice that $\tilde{V}^2_{eff}\to \tilde{E}^2$ as $\rho \to 2$
and $\rho \to 0$, and that $\tilde{V}^2_{eff}\to 1$ as $\rho \to
\infty$. Plots of (\ref{eq11})  for different values of
$\tilde{E}$ show a characteristic step--like behaviour in the
neighborhood of $\rho =2$ (see Figs. 1 and 2 in Ref. \cite{sch}).
An expansion of (\ref{eq11}) in the neighborhood of $\rho =2$
yields, in fact, $\tilde{V}^2_{eff}\sim\tilde{E}^2+\frac{(\rho
-2)^4}{4\epsilon^2\tilde{E}^4} +O((\rho -2)^5)$ which has the
minimum   $\tilde{E}^2$  on the horizon $\rho =2$. This term
vanishes only in the limit $\tilde{E}\to \infty$ and/or in the
limit $\epsilon\to \infty$, for which ${\cal A}_m$ or $M$ or both
vanish and the problem becomes meaningless.

The addition of MA effects does therefore produce a spherical shell of radius $2<\rho
<2+\eta$, with $\eta \ll 1$. The shell is classically impenetrable and remains so at
higher orders of approximation \cite{kerr}. The analogous occurrence of a classically
impenetrable shell was first derived by Gasperini as a consequence of the breaking of the
local $SO(3,1)$ symmetry \cite{MG}. A classically impenetrable shell and shift in horizon
also occur in the problem of particles in hyperbolic motion in a Kruskal plane \cite{8}.

The existence of large accelerations in proximity of the shell suggests the possibility
of radiation of photons (and gravitons) by bremsstrahlung. It has become recently clear
that production of electromagnetic radiation is possible even when the acceleration is
produced by a gravitational field\cite{boul,hig1,audr,hig2,crispino,crispino1}. Two
approaches are followed here.

$i)$ {\it Generalization of Larmor's formula.} A covariant generalization of this formula
would replace ${\ddot x}^\mu$ with its covariant counterpart which vanishes rigorously
along a geodesic in general relativity. It is of course know that even in Einstein's
theory a charged particle, while trying to adhere to the equivalence principle, does not
really follow a geodesic in a gravitational field. Deviations necessarily develop an
account of damping and of the {\it tail} effect discussed by De Witt and Brehme
\cite{dewitt}. In Caianiello's model, not only is the equivalence principle violated, but
the MA corrections also introduce deviations from geodetic motion. These are introduced
by $\sigma^2(x)$ which in first approximation is given by (\ref{eq6}). From the
definition (\ref{eq1}) of $\sigma^2(x)$ one finds
$g_{\mu\nu}\ddot{x^{\mu}}\ddot{x^{\nu}}=(\sigma^2-1){\cal A}_m^2$ and Larmor's formula,
which applies to particles of arbitrary trajectory, therefore becomes
\begin{equation}\label{larmor}
  P\simeq -\frac{2q^2}{3}\, {\cal A}_m^2(\sigma^2-1)\,.
\end{equation}
When the MA corrections vanish, $\sigma\to 1$ and $P\to 0$. Eq. (\ref{larmor}) therefore
seems suitable to deal with radiation due to MA. On the basis of (\ref{larmor}) and
(\ref{eq6}) one may expect radiation to be particularly intense when $\rho$ is close to
2, where $\sigma^2(x)$ has a singularity. One finds in fact
\begin{eqnarray}\label{6a}
 P& \simeq &
 -\frac{2q^2}{3}\, {\cal A}_m^2\,\frac{\epsilon^2}{\rho^7(\rho-2)^3}
 \left\{\lambda^4(\rho-2)^2(7-10\rho+3\rho^2)
   \right.  \nonumber \\
 & + & \left.   \rho^4[-4+4(1+2{\tilde E}^2)\rho+\rho^2(-1-4{\tilde E}^2+4{\tilde
 E}^4)] \right. \nonumber \\
 & -& \left. 2\lambda^2\rho^2(\rho-2)[10+(-19+10{\tilde E}^2)\rho+(11-8{\tilde
 E}^2)\rho^2+2({\tilde E}^2-1)\rho^3]\right\}.
 \end{eqnarray}
However, the particle can not reach the value $\rho=2$ because of
its acceleration and ensuing impenetrable shell. The shell's
radius corresponds to a maximum of ${\tilde V}_{eff}^2$ and to
the smallest value of $\rho$ the particle can reach. By expanding
${\tilde V}_{eff}^2$ as a series in the neighborhood of $\rho=2$
and keeping terms to sixth order, one finds
\begin{eqnarray}
 {\tilde V}_{eff}^2&\sim & \{
 256{\tilde
 E}^2(\rho-2)^4\epsilon^2+5(4+\lambda^2)^2(\rho-2)^6\epsilon^2+
 1024{\tilde E}^{10}\epsilon^4-16{\tilde E}^2(\rho-2)^5\times  \nonumber
 \\
 & &
 \times [-8(4+2\epsilon^2+\epsilon^2\lambda^2)+(16+4\epsilon^2+3\epsilon^2\lambda^2)\rho]\}
 \frac{1}{1024{\tilde E}^8\epsilon^4}
 \,, \label{pot-exp}
\end{eqnarray}
whose maximum is at
\begin{eqnarray}\label{7a}
 \rho_1&=&\{2(768{\tilde E}^2-240\epsilon^2+272{\tilde
 E}^2\epsilon^2-120\lambda^2\epsilon^2+164{\tilde
 E}^2\lambda^2\epsilon^2-15\lambda^4\epsilon^2) \\
 & + & 4{\tilde E}^2\epsilon\sqrt{6144{\tilde E}^2-1520\epsilon^2+1536{\tilde E}^2\epsilon^2
 -760\lambda^2\epsilon^2+1152{\tilde
 E}^2\lambda^2\epsilon^2-95\lambda^4\epsilon^2}\}\times \nonumber
 \\
 & & \times \{3(256{\tilde E}^2-80\epsilon^2+64{\tilde
 E}^2\epsilon^2-40\lambda^2\epsilon^2+48{\tilde
 E}^2\lambda^2\epsilon^2-5\lambda^4\epsilon^2)\}^{-1}\,. \nonumber
\end{eqnarray}
The distance of closest approach $\rho_1$ is real for
\[
{\tilde E}^2\geq
\epsilon^2(1520+760\lambda^2+95\lambda^4)(6144+153\epsilon^2+1152\lambda^2\epsilon^2)^{-1}
\]
and always positive for small values of $\lambda$ (radial infall) and $\epsilon$ which
are appropriate in many instances. It is useful to recall at this point the relationship
between the adimensional parameters $M$, ${\tilde E}$, $\lambda$, $\epsilon$, $\tau$
(adimensional time) and the corresponding (primed) CGS quantities. One has
 \[
 M=\frac{GM'}{c^2}\,, \quad {\tilde E}=\frac{E'}{m'c^2}\,, \quad
 \lambda=\frac{L'c}{GM'm'}\,, \quad \epsilon=\frac{\hbar
 c}{2GM'm'}\,, \quad \tau=\frac{c^3t'}{GM'}.
 \]
If $M'\sim 3M_\odot$ and the falling particle is a proton of energy $E'\sim 10$MeV, one
finds $\epsilon\sim 4.7\times 10^{-20}$ and ${\tilde E}\sim 0.01$. For these values of
the parameters $ 2-\rho_{1}\simeq 9.7\times 10^{-16}$.

An expression that gives $P$ as a function of $M$, ${\tilde E}$, $\lambda$ and $\epsilon$
can be obtained by substituting (\ref{7a}) into (\ref{6a}). Its usefulness is however
limited because the parameters involved can not yet be linked directly to the particle's
motion in the new field ${\tilde g}_{\mu\nu}$. This can be achieved by writing
$\displaystyle{\frac{dr}{d{\tilde s }}=\frac{dr}{dx^0}\frac{dx^0}{d{\tilde s}}}$, with
$\displaystyle{\frac{dx^0}{d{\tilde s}}=\frac{{\tilde E}}{{\tilde g}_{00}}=\frac{{\tilde
E}}{\sigma^2 g_{00}}}$, and using $\displaystyle{\left(\frac{dr}{d{\tilde
s}}\right)^2={\tilde E}^2-{\tilde V}^2_{eff}}$. One gets
\begin{equation}\label{8a}
 \left(\frac{dr}{dx^0}\right)^2=\frac{{\tilde E}^2-{\tilde V}^2_{eff}}{{\tilde
 E}^2}\sigma^4\left(1-\frac{2M}{r}\right)^2\,.
\end{equation}
The desired function $r=r(x^0)$ is obtained by integrating
(\ref{8a}) with respect to time. In adimensional variables, one
finds
\begin{equation}\label{9a}
  {\tilde E}\int \frac{\rho d\rho}{\sigma^2(\rho)(\rho-2)
  \sqrt{{\tilde E}^2-{\tilde
  V}^2_{eff}(\rho)}}=-\tau+a\,,
\end{equation}
where the negative sign in front of $\tau$ accounts for the fact
that the particle approaches the gravitational source as $\tau$
increases. The constant $a$ must be determined by appropriate
initial conditions.

\begin{figure}
\begin{center}
\resizebox{7cm}{!}{\includegraphics{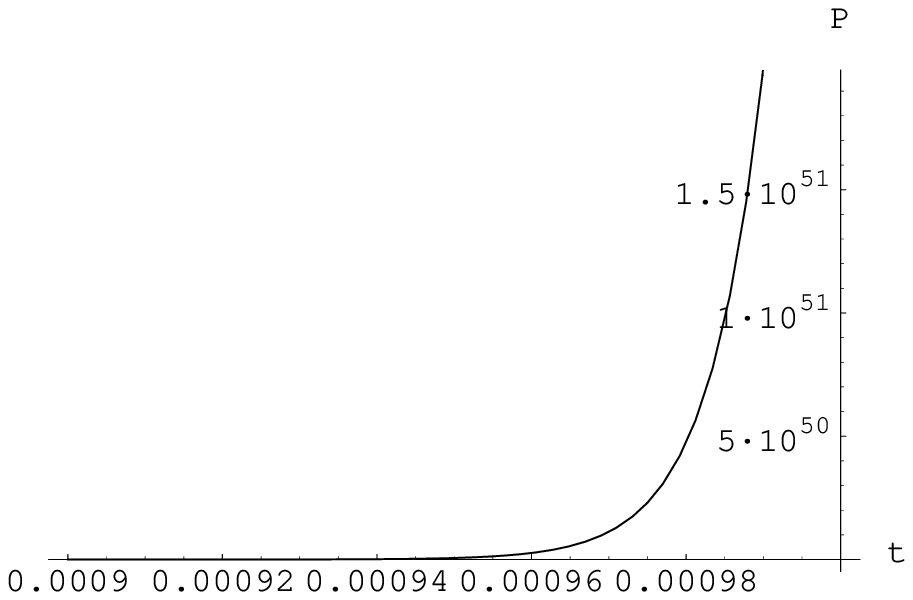}} \hfill
 \caption{\footnotesize {Power emitted as a function of time for $M'= 3 M_{\odot}, E'= 10 MeV,
 m'= 1.7\times 10^{-24}g$}}
 \label{Fig1}
\end{center}
\end{figure}

By assuming that the motion is radial ($\lambda=0$) and for the
values of ${\tilde E}$ and $\epsilon$ given above, the l.h.s. of
(\ref{9a}) can be integrated in the neighborhood of $\rho_1$ and
yields
 \begin{equation}\label{rho(t)}
 2\sqrt{2}\, \ln (\rho-\rho_1)\sim -\tau + a\,.
 \end{equation}
By requiring that $\rho\simeq 2.1$ at $t=0$, one finds $a=-1.3272$
and (\ref{rho(t)}) becomes
 \begin{equation}\label{10a}
 \rho\simeq \rho_{1}+ e^{-(\tau +6.5127)/2\sqrt{2})}\,.
 \end{equation}
The particle therefore reaches a distance within a $1\%$ of $\rho_{1}$, in the time
$t'\geq 4.8\times 10^{-5}$s. However, radiation becomes particularly intense when $t'\geq
1.68 \times 10^{-3}$. If (\ref{10a}) is then substituted into (\ref{6a}), one obtains
\begin{equation}\label{11a}
  |P|\sim \frac{2q^2{\cal A}_m^2}{3GM'}\, [2.67\times
  10^{42}+3.85\times 10^{47}(t'-1.68\times 10^{-3})+
  3.78\times 10^{52}(t'-1.68\times 10^{-3})^2]\,.
\end{equation}
Radiation becomes appreciable $\sim 10^{30}$erg/s only at $t'\geq 5\times 10^{-4}$s and
increases very rapidly with $t'$. As shown in Fig. \ref{Fig1}, it already is $P\sim
10^{51}$erg/s at $t'\sim 10^{-3}$s. Because of the quadratic dependence in both particle
charge and mass, small clumps of matter radiate significantly more and at higher
frequencies even a certain distance from $\rho_1$.

$ii)$ {\it Tree-level calculations.} The second way of estimating the power produced
follows the procedure of Crispino, Higuchi and Matsas \cite{crispino}. It treats the
electromagnetic field as a scalar field, whose source is the current produced by a
particle, also a scalar. Quantum field theory at the tree level is then used to describe
the electromagnetic field in a Schwarzschild background. The method is particularly
interesting in the context of Caianiello's model because it enables a comparison between
the massive scalar particle that produces the current and experiences the effective
geometry $\sigma^2 g_{\mu\nu}$, and the massless photon that can not be accelerated and
therefore "sees" only a Schwarzschild geometry. The source is the scalar current
\begin{equation}\label{current}
  j(x)=\frac{q}{\sqrt{-g}{\tilde
  u}^0}\delta(r-r_S)\delta(\theta-\pi/2)\delta(\phi)\,,
\end{equation}
normalized by the requirement that $\int d\sigma_V j(x)=q$, where $d\sigma_V$ is the
proper 3-volume element orthogonal to ${\tilde u}^\mu$, the four-velocity of the source.
The four-velocity component ${\tilde u}^0={\tilde E}\tilde{g}^{00}=\frac{{\tilde
E}}{\sigma^2 g_{00}}$ takes into account the MA corrections to the motion of the scalar
particle, whereas the 3-volume integration is in Schwarzschild space, hence the
$\sqrt{-g}$ in (\ref{current}). The scalar electromagnetic field generated by
(\ref{current}) is, at the tree level,
\begin{equation}\label{interact}
  A_{\omega l p}=i\int d^4x\sqrt{-g}\, j(x)u^*_{\omega
  lp}\,,
\end{equation}
where the functions
\begin{equation}\label{u(x)}
  u_{\omega lp}(x)=\sqrt{\frac{\omega}{\pi}}R_{\omega
  l}(r)Y_{lp}(\theta,\phi)e^{-i\omega t}
\end{equation}
represent the positive-frequency solutions ($\omega >0$) in Schwarzschild space-time of
the equation $\nabla_\mu\nabla^\mu u_{\omega lp}=0$. The functions $R_{\omega l}$ satisfy
the equation
\begin{equation}\label{wave-eq}
  e^{2\lambda}R_{\omega
  l}^{\prime\prime}+\left(\frac{2}{r}+\lambda^\prime\right)e^{2\lambda}R_{\omega
  l}^\prime+\left[\omega^2-e^\lambda\frac{l(l+1)}{r^2}\right]R_{\omega
  l}=0\,,
\end{equation}
where $e^\lambda=1-2M/r$ and the primes denote differentiation with respect to $r$
\cite{boson}. The emitted power is then \cite{crispino}
\begin{equation}\label{emittedP}
  P=\int_0^\infty d\omega \omega \frac{|A_{\omega l
  p}|^2}{T}\,,
\end{equation}
where $T$ is the time during which the interaction is switched on\cite{itzy}. Introducing
(\ref{current}) into (\ref{interact}) and using the coordinate $\rho-2=x$, one finds, to
leading order,
\begin{equation}\label{interact-calc}
  A_{\omega l p}\simeq \frac{De^{6.5127/\sqrt{2}}}{2M\sqrt{\omega}}
  \frac{e^{(iM\omega+1/\sqrt{2})\tau}-1}{iM\omega+\frac{1}{\sqrt{2}}}\,,
\end{equation}
where
 \[
 D\equiv i\, \frac{4q{\tilde E}^3}{{\cal
 A}_m^2M\sqrt{\pi}}\,Y_{lp}(\pi/2, 0)(1+\tilde{R}_{\omega l})\,.
 \]
The contribution of the term $-1$ on the r.h.s. of (\ref{interact-calc}) is negligible
relative to the exponential. The reflection coefficient $\tilde{R}_{\omega l}<1$ is also
neglected.

\begin{figure}
\begin{center}
\resizebox{7cm}{!}{\includegraphics{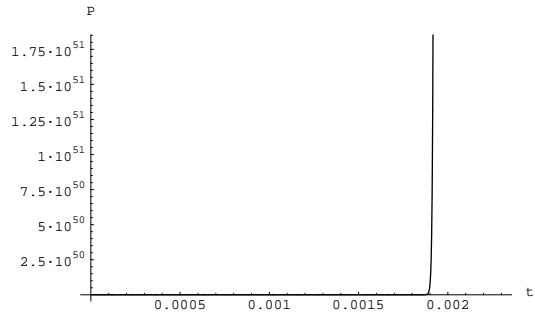}} \hfill \caption{ \footnotesize {Power
emitted as a function of time according to Eq.(\ref{power-rad-cgs}). The values of $M',
E', m'$ are as in Fig.1}} \label{Fig2}
\end{center}
\end{figure}

\begin{figure}
\begin{center}
\resizebox{7cm}{!}{\includegraphics{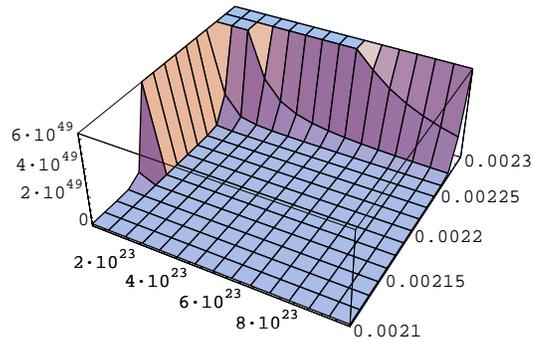}} \hfill \caption{\footnotesize {Power
spectrum in the $\gamma$-ray frequency region and the time interval $2.1 ms$ to $2.3 ms$.
The values of $M', E', m'$ are as in Fig.1}} \label{Fig3}
\end{center}
\end{figure}

Eqs. (\ref{emittedP}) and (\ref{interact-calc}) give the radiated
power as
\begin{equation}\label{power-rad}
  |P|\simeq \frac{q^2{\tilde E}^6e^{6.512}}{\pi {\cal
  A}_m^4M^6}\frac{e^{\sqrt{2}\,\tau}}{\tau},
\end{equation}
where the time $T=M\tau^\prime$ in (\ref{emittedP}) and $\tau^\prime$ is the time
required for the particle to reach the intense radiation zone of its trajectory. Though
this result vanishes as $\hbar\to 0$, the correct limit for vanishing MA contributions
follows by taking the limit $\sigma\to 1$ in (\ref{current}) and is finite. In CGS units,
the power becomes
\begin{equation}\label{power-rad-cgs}
  |P|\simeq \frac{q^2E^{\prime 6} \hbar^4 e^{8.21}}{16 \pi m' G^5M^{\prime 5}}
  \frac{e^{\sqrt{2}\,c^3t'/GM'}}{t'}
  \simeq 2.35\times 10^{-110} \frac{e^{1.9\times 10^5 t'}}{t'}\,.
\end{equation}
Despite the enormously small reducing factor in (\ref{power-rad-cgs}), a single proton
reaches powers of $\sim 10^{30}$erg/s after only $t'\sim 2\times 10^{-3}$s, and $P\sim
10^{51}$erg/s is reached for $t'\sim 2.3\times 10^{-3}$s, in agreement with the results
of (\ref{11a}). The frequency distribution of $P$ is given by
\begin{equation}\label{freq-distrib}
  \frac{dP}{d\omega}\simeq
  \frac{|D|^2e^{9.121}}{4M^3}\,\frac{e^{\sqrt{2}\,
  \tau}}{(M^2\omega^2+\frac{1}{2})\tau}\,,
\end{equation}
or, in CGS units, by
\begin{equation}\label{freq-distrib-cgs}
  \frac{dP}{d\omega}\simeq
  \frac{10^4}{4\pi^2}\,
  \frac{\hbar^4 E^{\prime 6} q^2}{G^4 M^{\prime 4} m^{\prime 10} c^8}\,
  \frac{e^{\sqrt{2}\,c^3 t'/GM'}}{(\frac{G^2 M^{\prime 2} \omega^2}{c^6}+\frac{1}{2})t'}\,.
\end{equation}
It can be seen from the Figs. \ref{Fig2} and \ref{Fig3} that the bursts of radiation are
sudden and peaked at lower frequencies, though the intensity of radiation also remains
high at $\gamma$-ray frequencies. Some similarities with $\gamma$-ray busters are
obvious. The latter have extraordinary large energy outputs $\sim 10^{51}$ to $\sim
10^{54}$erg/s. Their spectra are non-thermal with photon energies ranging from $50keV$ to
the $GeV$ region. In CIAO's case spectra also are non-thermal, in fact typically of the
bremsstrahlung type, with intense radiation that extends to the same frequency range of
bursters. Significant amounts of low frequency radiation could be absorbed by the
progenitor, if this is a black hole, in the case of busters. Black holes are known to
have sizeable absorption for infrared photons \cite{crispino}. For CIAOs absorption could
also take place at the shell which, for photons, behaves like a horizon. Nonetheless
energy extraction from CIAOs remains extremely efficient. The duration of the intense
emission by particles in their proximity is of the same order of magnitude, $1ms$ to
$10^3$s, of the events that are observed. Finally, the lower frequency afterglows that
are sometimes observed to last several months, may be explained, in CIAO's scenario, as
due to a dip in the effective potential experienced by scalar particles at higher
accelerations\cite{boson}. The material trapped in this way may subsequently leave the
relatively shallow attractive area to resume the acceleration and radiation processes at
lower intensities and frequencies.

\newpage

\bigskip
\bigskip
\begin{centerline}
{\bf Acknowledgments}
\end{centerline}
Research supported by MURST PRIN 2001 and by the Natural Sciences
and Engineering Research Council of Canada.

\newpage

\end{document}